\documentclass[10pt,aps,prx,twocolumn,superscriptaddress,nofootinbib,floatfix,amssymb,showpacs]{revtex4-2}
\usepackage[english]{babel}

\makeatletter
\def\l@en{\l@english}
\makeatother

\usepackage[skip=0 pt, indent=10pt]{parskip}
\usepackage{graphicx}
\usepackage{times}
\usepackage{amsmath}
\usepackage{nccmath}
\usepackage{soul}
\usepackage{color}
\usepackage{comment}

\usepackage[colorlinks=false,citecolor=blue,urlcolor=blue,bookmarks=false,hypertexnames=true]{hyperref} 
\usepackage{url}
\usepackage{cleveref}

\begin{document}



\title{Ultra-small Mode Volume Polariton Condensation via Precision $He^+$ Ion Implantation}



\author{Y. C. Balas}
\email{i.balas@westlake.edu.cn}
\affiliation{Department of Physics, School of Science, Westlake University, Hangzhou 310014, China}

\author{X. Zhou}
\affiliation{Department of Physics, School of Science, Westlake University, Hangzhou 310014, China}

\author{E. Cherotchenko}
\affiliation{Ioffe Institute, ul. Polytekhnicheskaya 26, St. Petersburg, 194021, Russian Federation}

\author{I. Kuznetsov}
\affiliation{Moscow Institute of Physics and Technology, Institutskiy per., 9, Dolgoprudnyi, Moscow Region 141701, Russia}

\author{S.K. Rajendran}

\affiliation{SUPA, University of St. Andrews, St. Andrews KY16 9SS, United Kingdom}

\author{G.G. Paschos}
\affiliation{Department of Physics, School of Science, Westlake University, Hangzhou 310014, China}

\author{A.V. Trifonov}
\affiliation{Spin Optics Laboratory, St. Petersburg State University, Ulyanovskaya 1, St. Petersburg 198504, Russia}

\author{A. Nalitov}
\affiliation{Moscow Institute of Physics and Technology, Institutskiy per., 9, Dolgoprudnyi, Moscow Region 141701, Russia}
\affiliation{Russian Quantum Center, Skolkovo, Moscow, 121205, Russia}

\author{H. Ohadi}

\affiliation{SUPA, University of St Andrews, St Andrews KY16 9SS, United Kingdom}

\author{P.G. Savvidis}
\email{p.savvidis@westlake.edu.cn}
\affiliation{Department of Physics, School of Science, Westlake University, Hangzhou 310014, China}
\affiliation{Institute of Electronic Structure and Laser and Center for Quantum Science and Technologies, FORTH, 70013 Heraklion, Greece}

\date{\today}

\begin{abstract}
We present a novel method for generating potential landscapes in GaAs microcavities through focused $He^{+}$ implantation. The ion beam imprints micron-scale patterns of non-radiative centers that deplete the exciton reservoir and form a loss-defined potential minimum. Under non-resonant pumping, the resulting traps have a lateral size $\le 1.2 ~\mathrm{\mu m}$ and a three-dimensional mode volume of only $\approx 0.6 ~ \mathrm{\mu m^3}$, small enough to to support a single polariton condensate mode. The implantation process maintains strong coupling and provides lithographic ($ < 300 ~ \mathrm{nm}$) resolution. These loss-engineered traps effectively overcome the micrometer-scale limitations of conventional microcavity patterning techniques, opening new avenues for device development and polariton research within the quantum regime.
\end{abstract}


\maketitle

\section{Introduction}

Exciton-polaritons are hybrid light-matter quasiparticles formed from the strong coupling of photons and excitons in microcavities. As light-matter quasiparticles, they possess an ultrasmall effective mass, enabling condensation even at room temperature, while also exhibiting significant polariton-matter interactions \cite{kasprzak_boseeinstein_2006}. Tailoring the in-plane potential of the microcavity (MC) is the workhorse of polaritonic studies. This approach facilitates the exploration of phenomena such as non-equilibrium quantum fluids \cite{carusotto_quantum_2013}, topological photonics \cite{klembt_exciton-polariton_2018}, ultrafast optoelectronics \cite{ballarini_all-optical_2013}, and low-dimensional quantum phenomena \cite{moreau2001single,santori2002indistinguishable,heindel2010electrically,klaas_photon-number-resolved_2018,bajoni_polariton_2008}. A broad set of techniques is now employed to shape microcavity potentials, advancing both fundamental research and practical device development.

Taking advantage of the fact that polariton condensates are realized in semiconductor microcavities, the most established methods for achieving tight confinement are provided by semiconductor micro- and nanofabrication techniques. Etched $GaAs/AlGaAs$ micropillars exhibit lasing at diameters as small as $1 ~\mu m$, reaching single-mode volumes near $10 ~\mu m^3$ \cite{czopak_polariton_2021}. Open-access microcavities confine polaritons without the need for etching and allow for in-situ tuning of the Rabi splitting \cite{li_condensation_2022}. Photonic crystal defect cavities and bound-state-in-continuum lattices have further enhanced confinement to the wavelength scale, producing room-temperature condensates \cite{wu_exciton_2024}. Because the interaction-induced blueshift scales inversely with mode volume, shrinking the mode volume enhances polariton–polariton nonlinearities in the few-particle regime, where phenomena such as polariton blockade are predicted \cite{PhysRevB.73.193306}. Although these approaches have reduced condensate sizes to a few microns, fabrication disorder in etched pillars, and intrinsic absorption continue to limit coherence, quality factor, and scalability.

Optically imprinted annular potentials have produced trapped condensates that are delocalized from the excitonic reservoir \cite{askitopoulos_polariton_2013}. Rotating superfluids and deterministic vortex clusters \cite{del_valle-inclan_redondo_optically_2023}, as well as, most recently, vortex "qubit" analogues \cite{barrat_qubit_2024}. However, the diffusion of the long-lived exciton reservoir blurs these additive potentials over several microns, setting a resolution limit.

Focused ion beam implantation provides an alternative route that combines precision and repeatability with the relative ease of potential imprinting. Sub-micron $He^+$ stripes in GaAs quantum wells introduce controlled non-radiative broadening while preserving the integrity of cavity optics \cite{kapitonov_effect_2015}. High-energy proton doses can independently tune exciton and photon energies by intermixing the quantum wells and the distributed Bragg reflector (DBR) layers near the interfaces \cite{fraser_independent_2023}.

Here, we demonstrate $He^+$ implanted traps with lateral dimensions of $\leq 1.2 ~\mu m$ and mode volumes of $< 1 ~\mu m^3$, enabling condensation with fewer than 200 polaritons. The implantation process locally accelerates non-radiative decay while maintaining strong coupling. The non-radiative centers drain the exciton reservoir, thereby reducing the repulsive blueshift potential. Under pumping, the depleted reservoir forms a trapping potential. These traps feature excellent uniformity and a resolution that surpasses the limitations of both lithographic and optically written patterns. This platform facilitates the implementation of complex design patterns that incorporate gradient potentials with nanoscale precision, opening a path toward few-particle polariton devices and vortex-based qubits.

\section{$He^+$  Ion Implantation}

\begin{figure}[tb]
\centerline{\includegraphics[width=\columnwidth]{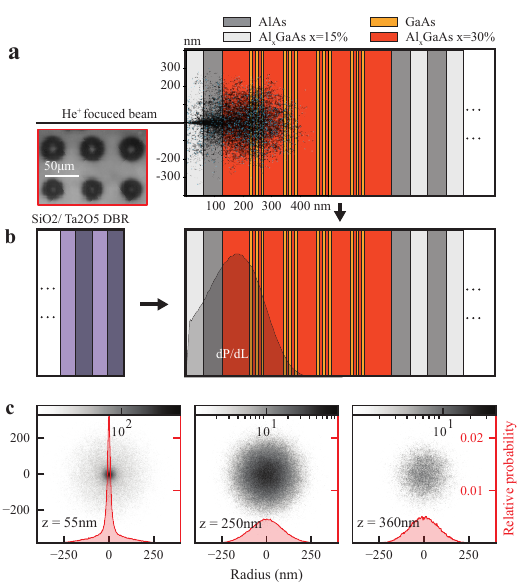}}
\caption{Schematic representation of sample fabrication steps. (a) Step one: Simulated diffusion of vacancies (black dots) caused by $He^+$ ion implantation (blue dots). The insert in red frame is a PL image of the sample without the top DBR; the dark (black) regions were implanted. (b) Post implantation deposition of top dielectric DBR. Shaded area: Relative density of structural defect versus sample depth. (c) Lateral diffusion length of defects for different depth zones of 50 nm; Central depths denoted in each frame.}
\label{fig:implantation}
\end{figure}

A $35~keV$ Focused $He^+$ ion beams is utilized to irradiate the half-microcavity samples at doses ranging from $1 \times 10^{11}~\mathrm{ions/cm^2}$ to $1 \times 10^{12}~\mathrm{ions/cm^2}$. This process introduces structural defects into the crystal lattice, primarily generating vacancies (indicated by black dots \ref{fig:implantation}a) with smaller densities of interstitial and substitutional defects, which are marked with blue in \ref{fig:implantation}a. TRIM ion scattering simulations \cite{ziegler_srim_2010} indicate that approximately 180 vacancies are generated per implanted ion. To maintain a high depth-to-lateral aspect ratio of the ion implantation density, we resort to near-surface implantation of a microcavity structure without the top dielectric DBR mirror and deposit top DBR post-implantation.

The sample under investigation consists of a half GaAs/AlGasAs-based semiconductor heterostructure grown by molecular beam epitaxy (MBE), and a top dielectric distributed Bragg reflector (DBR) fabricated through electron beam evaporation. Four sets of three 12 nm GaAs/AlGaAs quantum wells (QWs) are embedded between 35 pairs of AlGaAs/AlAs bottom DBR and a top $\mathrm{Ta_2O_5/SiO_2}$ dielectric DBR. To maintain a high depth-to-lateral aspect ratio of the ion implantation density, we employ near-surface implantation of the incomplete microcavity, prior to the deposition of the top DBR (Fig.~\ref{fig:implantation}b).

Simulations of the experimental conditions used for ion implantation reveal good overlap in the peak defect density at the depth of the first set of quantum wells (QWs), which ranges from 220 to 270 nm from the surface (Fig.~\ref{fig:implantation}b). The lateral defect span at this depth is approximately 300 nm (Fig. ~\ref{fig:implantation}c, middle). The current implantation technique allows for $<50~nm$ resolution for QWs located near the surface (Fig.~\ref{fig:implantation}c, left). However, the enhancement of Rabi splitting from multiple surviving QWs is also desirable.

\section{QW optical characterization}
\begin{figure}[tb]
\centerline{\includegraphics[width=\columnwidth]{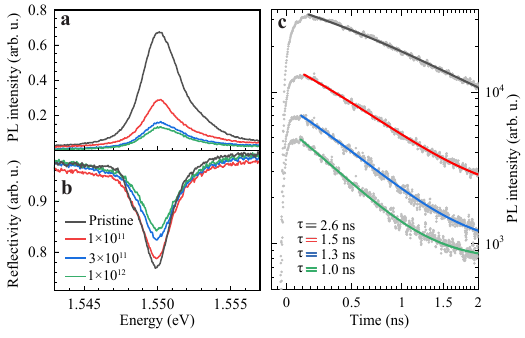}}
\caption{(a) Pl quenching, (b) reduction of resonant QW absorption, and (c) reduction of PL emission lifetime on various ion implantation doses in QWs. }
\label{fig:excitons}
\end{figure}

Photoluminescence (PL), Time-Resolved PL (TRPL), and reflectivity measurements at 4 K are used to probe the effects of $He^+$ implantation on the optical response of half-microcavity QWs. Micro-reflectivity spectra are obtained using a broadband white-light source, which is directed onto the sample through a microscope objective. Implantation-induced vacancies create non-radiative recombination centers, reducing resonant excitonic absorption as the implantation dose increases (Fig.~\ref{fig:excitons}b).

A continuous-wave Ti:Sapphire laser is used to optically pump the sample non-resonantly at $1.710~ \mathrm{eV}$ with $\approx 1~\mathrm{\mu m}$ spot while emitted PL is collected through a microscope objective. A long-pass filter cuts off the reflection of the pump laser. The PL intensity decreases with increasing helium ion doses due to the enhancement of non-radiative pathways, which reduces the quantum efficiency of the QWs (Fig.~\ref{fig:excitons}a).

For TRPL experiments, we utilize a mode-locked $150~\mathrm{fs}$ pulse Ti:Sapphire laser, and PL decay dynamics is measured using a streak camera across a range of implantation doses. We observe a clear reduction in PL lifetime ($\tau$) with increasing ion doses. (Fig.~\ref{fig:excitons}c), a clear signature of enhanced non-radiative recombination rates.

Several conclusions can be drawn from these measurements based on the rate equation, governing the dynamics of exciton density $N$:
\begin{equation}
    {dN \over dt} = P - (\gamma_R+\gamma_{NR})N,
\end{equation}

where $P$ is the pump term and $\gamma_{\mathrm R}$ ($\gamma_{\mathrm{NR}}$) are the radiative (non-radiative) rates.  Implantation increases $\gamma_{\mathrm{NR}}$, which (i) lowers the steady-state exciton density $N=P/(\gamma_{\mathrm R}+\gamma_{\mathrm{NR}})$ and therefore the blueshift potential and (ii) reduces the PL yield $I=N\gamma_{\mathrm R}$ without noticeably altering $\gamma_{\mathrm R}$.  The resulting spatial modulation of $N$ (high outside, strongly depleted inside the implanted area) acts as an effective non-Hermitian potential, a mechanism we exploit in the confinement experiments that follow.

\section{Polariton condensate confinement}

We continue our investigation of the DBR-capped MC. By increasing the pumping power of the pump laser, we compare the polariton condensation thresholds between the pristine and implanted regions (see Appendix section \ref{sec:threshold}). Ion implantation primarily affects the first set of 3 QWs leaving $\approx75\%$ of the oscillator strength intact. Since the vacuum-Rabi splitting scales as $\Omega_R\propto\sqrt{N_{\text{osc}}}$, the maximum expected reduction is $\sqrt{0.75}=0.87$. Notably, under front-side excitation used in the experiments, a large fraction of carriers is generated within the topmost QWs that are strongly affected by the rapid non-radiative recombination in these QWs. Consequently, we observe condensation in all regions: pristine and implanted, with their respective condensation threshold powers significantly increasing with higher implantation doses.

\begin{figure}[tb]
\centerline{\includegraphics[width=\columnwidth]{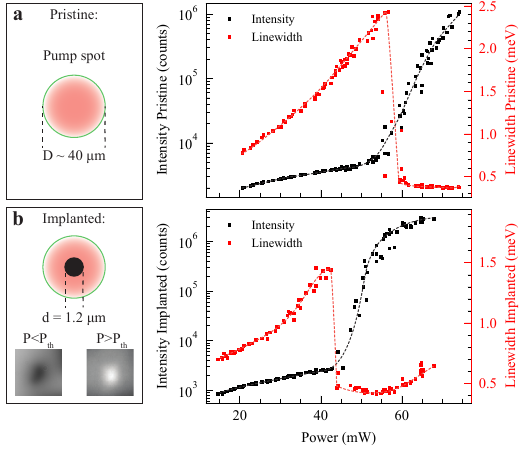}}
\caption{Power dependence (pump power vs emitted PL intensity) from a pristine region of the sample (a), and from an area of the sample with a $d=1.2\mu m$ implanted spot (b). Inset images show that the source of PL emission under threshold (un-implanted region) and over threshold (implanted region) . Both measurements where performed using a $D \approx 40 \mu m $ laser spot. }
\label{fig:trap}
\end{figure}

Subsequently, the excitation beam was expanded to $40\mu m$, surrounding an implanted dot measuring $1.2 \mu m$ in diameter (Fig. \ref{fig:trap}a,b (left)). Since the blueshift potential depends on exciton density, the implanted dot effectively forms a potential well due to enhanced excitonic losses. Polaritons condensed within the implanted dot exhibit clear spatial confinement (Fig. \ref{fig:trap}b (left) insets). Remarkably, this configuration has lower condensation threshold (Fig.\ref{fig:trap}b (right)) compared to direct excitation without the implanted region (Fig. \ref{fig:trap}a (right)). This trapping scheme highlights the effectiveness of engineered defect loss landscapes in reducing exciton density, offering a novel pathway for achieving efficient polariton confinement and condensation. This tight confinement leads to a reduction in the steady-state number of polaritons $N$. Specifically, $N_{tr} \approx 150$ while $N_{pr} \approx 1300$, where $N_{tr}$ corresponds to the trapped condensate shown in figure \ref{fig:trap}, and $N_{pr}$ corresponds to the pristine sample condensation at $P= 1.6P_{th}$. For details regarding the polariton number calculation see Appendix \ref{sec:A2}.

\begin{figure}[tb]
\centerline{\includegraphics[width=\columnwidth]{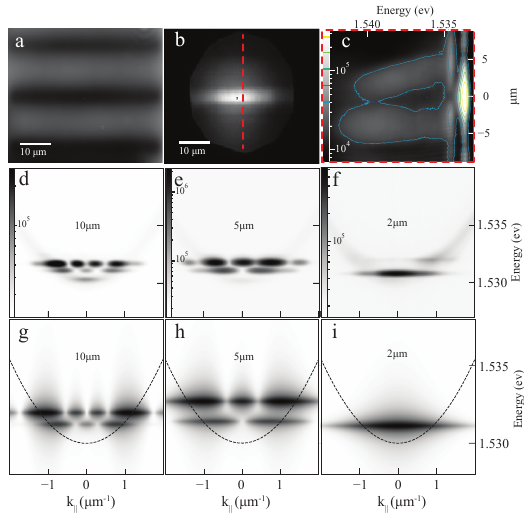}}
\caption{Top row: Real Space image of a region implanted with $W = 5 \mu m$ wires (dark regions) under the condensation threshold (a), above condensation threshold (b). (c) is the spectral analysis of the red cross section in (b). Middle row: Far field spectroscopy above the condensation threshold of wires with widths $W_d = 10 \mu m $, $W_e = 5 \mu m $, $W_f = 2 \mu m $. Bottom row: Analytical calculation (g,h,i) of (d,e,f) respectively.
Parameters: $l_0 = 3\,\mu$m, $\varepsilon = 2.5$, $n_0 = 7.5$, $c=0.1$.}
\label{fig:wires}
\end{figure}

The universality of the trapping technique is tested with patterning microwires of various widths (Fig.~\ref{fig:wires}a), demonstrating confinement effects analogous to those observed in a quasi-1D quantum well (QW). Real-space imaging revealed bright emission from the center of the wire for $P > P_{th}$ (Fig.~\ref{fig:wires}b). An energy-resolved vertical slice taken from the center of figure \ref{fig:wires}b reveals that the emission from the edges of the implanted wire exhibits suppressed intensity and higher energy compared to the center of the trap, which emits brightly at lower energy (Figure \ref{fig:wires} c). Angle-resolved emission indicates that the energy levels depend on wire width, consistent with theoretical predictions for 1D potential wells (Fig.~\ref{fig:wires}d-f).

We describe the polariton condensates in quasi-1D wires near the condensation threshold, neglecting nonlinearities, with the non-Hermitian variation of 1D Schroedinger equation:

\begin{equation} 
    \label{eq:1DShr}
    \left[-{\hbar^2 \over 2 m} \partial_x^2 + {\alpha+i\beta \over 2}N(x) - i{\hbar \Gamma \over 2 } -i\hbar \partial_t \right] \Psi(x,t) = 0,
\end{equation}
where $m$ is the effective polariton mass, $N(x)$ is the spatially profiled exciton reservoir density, $\alpha$ and $\beta$ are the parameters governing the strength of conservative and dissipative parts of interaction with the reservoir, and $\Gamma$ is the polariton radiative loss rate.

We then assume a sharp shape of the reservoir density profile: $N(x)=0$ if $|x|<D/2$ and $N_0$ if $|x|>D/2$, with $D$ the wire width.
The linear spectral properties of the Eq. \ref{eq:1DShr} are then obtained by solving the non-Hermitian variation of the textbook 1D QW problem.

We first introduce characteristic scales in time and space $t_0 = 2/\Gamma$ and $l_0 = \sqrt{\hbar/(2m\Gamma)}$ to obtain the dimensionless stationary variant of Eq. \ref{eq:1DShr}:
\begin{equation}
    \tilde{\omega} \Psi = \left[ -\partial_\chi^2 + (\varepsilon + i)n - i \right] \Psi,
\end{equation}
where $\chi = x/l_0$, $\varepsilon = \alpha/\beta$, and $n=2\beta N/(\hbar \Gamma)$.

In full analogy with the conservative problem, the effective Hamiltonian symmetry allows even and odd eigenstates $\Psi(\chi)=\pm\Psi(-\chi)$.
Outside the doped region ($|\chi|>d/2$) with $d=D/l_0$, where the reservoir density takes the value $n_0=2\beta N_ 0(\hbar \Gamma)$, it is convenient to denote the solution vanishing at $|\chi|\rightarrow{\infty}$ as $B\exp[-\varkappa (\chi-d/2)]$ for $\chi>d/2$ and $\pm B\exp[\varkappa (\chi-d/2)]$ for even and odd states at $\chi<-d/2$.
Here $\varkappa=\sqrt{(\varepsilon+i)n_0-i-\tilde{\omega}}$ and the principal branch $\mathrm{Re}\lbrace \varkappa \rbrace > 0$ is selected to avoid divergence at $|\chi|\rightarrow{\infty}$.
Notably, such evanescent solutions simultaneously satisfy the boundary conditions for polariton currents $J(\chi) = \hbar\mathrm{Im}\lbrace \Psi^*\partial_\chi\Psi\rbrace/m$, namely, $J(\chi>d/2)>0$ and $J(\chi<-d/2)<0$.
In the ion-doped region, the reservoir density is reduced by a factor $c<1$, producing an effective potential $(\varepsilon+i)cn_0$.
The corresponding even and odd wave functions read $\Psi = A\cos(q\chi)$ and $\Psi = A\sin(q\chi)$, respectively, where $q = \sqrt{\tilde{\omega}+i-(\varepsilon+i)cn_0}$ with the selection of the principal square root branch $\mathrm{Re}\lbrace q \rbrace>0$.

The eigenstates are obtained by matching the piecewise defined wave functions at $|\chi|=d/2$, or equating $\tan(qd/2)$ to $\varkappa/q$ for even states and to $-q/\varkappa$ for odd states.
Solving for $\tilde{\omega}$ on the complex plane yields the spectrum of discrete trapped polariton modes $\tilde{\omega}_j$ along with the corresponding wave functions $\Psi_j(x)$.
While the emission energy of a polariton mode is given by the real part of its energy $\omega_j = \mathrm{Re}\lbrace \tilde{\omega}_j \rbrace \in \left[ c \varepsilon_0, \varepsilon n_0\right]$, the imaginary part $g_j = \mathrm{Im}\lbrace \tilde{\omega}_j \rbrace \in \left[ -1 + cn_0, -1+n_0\right]$ describes its gain rate and, at the same time, the stationary population under cw pumping, provided $g_j>0$.

The reciprocal space emission profiles $|\tilde\Psi_j(k)|^2$ of the trapped polariton modes can be analytically expressed in the energies $\tilde\omega_j$ by calculating the Fourier transform of $\Psi_j(x)$, see Appendix \ref{AppA}.
The total multimode emission profile along the direction, perpendicular to the microwire, is given by the sum of all contributions over all trapped modes, characterized with stationary populations $g_j>0$ and unitary broadening, given by the Lorentzian factor:
\begin{equation} \label{eq:emission}
    I(k,\omega) = \sum_j {g_j |\tilde{\Psi}_j(k)|^2 \over 1+(\omega-\omega_j)^2},
\end{equation}

\section{Ring traps}

\begin{figure}[tb]
\centerline{\includegraphics[width=\columnwidth]{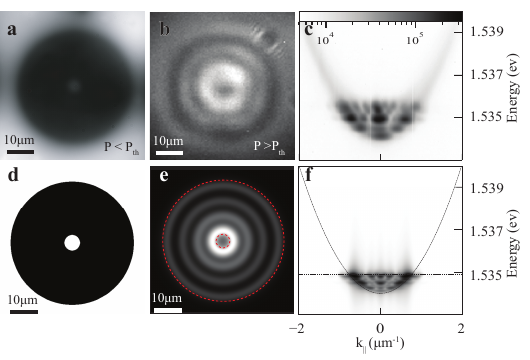}}
\caption{(a) Real space of PL emission below threshold $(P < P_{\text{th}})$: the implanted region appears dark. (b) above threshold $(P > P_{\text{th}})$: a bright set of concentric rings appears inside the implanted annulus, evidencing quantised radial modes of the condensate. (c) Angle-resolved emission at  $(P > P_{\text{th}})$ (logarithmic colour scale) showing discreet energy states. (d) Binary loss mask used in the numerical model (black = high loss). (e) Steady-state condensate density obtained from the non-Hermitian Schrödinger solutions are in agreement with experimental results; the red dashed circle marks the implanted rim. (f) Simulated k-space spectrum  of confined modes. The dashed curve represents the un-confined polariton lower polariton branch. Resulting potential depth marked with a dashed horizontal line.
Parameters: $l_0=5\,\mu$m, $\varepsilon = 5$, $n_0=4.5$, $c=0.25$.}
\label{fig:ring}
\end{figure}

In ring-shaped traps with outer diameters of $D = 35 \mu$m and inner diameters of $d = 4 \mu$m (Fig \ref{fig:ring}a), the excitation beam generates a radial potential profile. The unimplanted center accumulates a high exciton density under the excitation beam, resulting in a pronounced blueshift potential spike. The surrounding implanted area, characterized by reduced exciton density, forms a annular potential well. Polaritons condense within this well, exhibiting radially symmetric modes (Fig \ref{fig:ring}b). Angle-resolved emission confirms the presence of discrete energy levels associated with this confinement (Fig \ref{fig:ring}c).

Similarly to the case of microwires, non-Hermitian polariton confinement in isotropic ring-shaped traps is described by the equation
\begin{equation} \label{eq:2DShr}
    \left[ -{\hbar^2 \over 2 m} \Delta_{r,\varphi} + {\alpha+i\beta \over 2}N(r) - i{\hbar \Gamma \over 2 } - i \hbar \partial_t \right] \Psi(r,\varphi,t) = 0,
\end{equation}
with $\Delta_{r,\varphi}$ being the Laplacian expressed in polar coordinates.

Sharp-edge annular traps correspond to a piecewise-defined reservoir density: $N(r)=N_0$ if $r<d/2$ or $r>D/2$, $N(r) = c N_0$ if $d/2<r<D/2$.

The dimensionless stationary variant of Eq. \eqref{eq:1DShr} reads
\begin{equation}
    \left[ -{1 \over \rho } {\partial \over \partial \rho} \left( \rho {\partial \over \partial \rho} \right) - {1\over \rho^2} {\partial^2 \over \partial \varphi^2}   + (\varepsilon + i)n - i - \omega \right] \Psi(\rho,\varphi),
\end{equation}
where $\rho = r/l_0$.

The radial symmetry of the effective potential, stemming from the reservoir, and the boundary conditions at $r=0$ and $r \rightarrow \infty$ govern the form of the wave function $\Psi_l(r,\varphi) = \Phi_{l}(\varphi)R_{l}(r)$:
\begin{equation} \label{eq:WF2Dreal}
    \Psi_l(\rho,\varphi) = \exp(il\varphi) \times  \begin{cases}
    A I_l(\varkappa\rho), & \rho<\rho_1, \\
    B J_l(q\rho) + C Y_l(q\rho), &  \rho_1<\rho<\rho_2, \\
    D K_l(\varkappa\rho), & \rho>\rho_2 \
    \end{cases} 
\end{equation}
where $A,B,C,D$ are complex coefficients related with the normalization condition $2\pi\int|\Psi(r,\varphi)|^2 rdr=1$.

Matching the wave function values and spatial derivatives at $\rho_1 = R_1/l_0$ and $\rho_2 = R_2/l_0$ yields the transcendental complex equation for complex energies $\tilde\omega_{l,j}$ of the effectively trapped polariton states, labeled with integer angular and radial numbers $l$ and $j$.
The analytical expressions for the Fourier transforms of the wave functions $\tilde\Psi_{l,j}(k,\varphi_k)$ is given in Appendix \ref{AppA}.
Due to the radial trap symmetry, the emission spectrum $I(k,\omega)$ is isotropic and is determined by the same expression as in Eq. \eqref{eq:emission}, but with summation carried out over all integer $l$ and $j$, corresponding to macroscopically populated polariton modes $g_{l,j}=\mathrm{Im}\lbrace \tilde\omega_{l,j}\rbrace>0$, emitting at frequencies $\omega_{l,j}=\mathrm{Re}\lbrace \tilde\omega_{l,j}\rbrace$.

\section{Conclusion}
Focused $He^+$ ion implantation provides a robust technique for engineering lateral confinement in polariton microcavities. By tailoring the patterns we achieved high-precision control over polariton condensates in dots, microwires and ring-shaped traps. These findings emphasize the role of excitonic loss landscapes in the design of efficient polariton traps and highlight the potential of this technique for scalable quantum photonic applications.

\renewcommand{\thefigure}{A.\arabic{figure}}
\setcounter{figure}{0}

\appendix
\section{Condensation threshold versus implantation dose} 
\label{sec:threshold}
\begin{figure}[ht]
\centerline{\includegraphics[scale = 1]{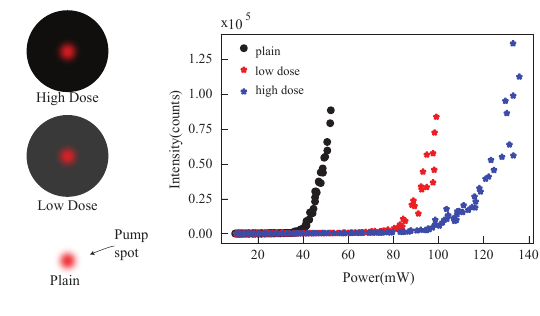}}
\caption{(Left): Schematic of the pumping scheme; the red dot represents the laser pump. The dark regions represent the implanted regions. (Right) Power dependence of intensity emitted in the pristine (plain) (black circles) and the regions implanted with $1{\times}10^{11}~ \text{ions} \cdot \text{cm} ^{-2}$ (red stars), and $1{\times}10^{12}~ \text{ions} \cdot \text{cm} ^{-2}$ (blue stars). The extracted thresholds are $\approx 40~ \text{mw}$, $\approx 80~ \text{mw}$, and $\approx 110~ \text{mw}$ respectively.}
\label{fig:thresholds}
\end{figure}

Figure \ref{fig:thresholds} characterizes the effect of implantation \emph{alone}, without the lateral trap geometry of the main Letter. A tightly focused ($\sim1\mu$m) non-resonant beam was positioned on three different regions of the complete microcavity: pristine (“plain”), low-dose He$^{+}$ implantation ($1{\times}10^{11}~ \text{ions} \cdot \text{cm} ^{-2}$), and high-dose implantation ($1{\times}10^{12}~ \text{ions} \cdot \text{cm} ^{-2}$). In this configuration, the potential landscape is essentially flat; implantation modifies only the local exciton lifetime. Therefore, more pump power is required to accumulate the critical polariton density. Specifically, the thresholds range from $\approx 40~ \text{mw}$ (plain) to $\approx 80~ \text{mw}$ (low dose), and $\approx 110~ \text{mw}$ (high dose).

\renewcommand{\theequation}{B.\arabic{equation}}
\setcounter{equation}{0}
\section{Polariton number derivation} \label{sec:A2}

We estimate the polartion number at $P = 1.6P_{th}$ in the $1.2~\mu m$ trap according to the technique described in \cite{PhysRevX.5.031002}. Polariton number is given by:
\begin{equation}
    N = \frac{\alpha R \tau}{n_{tot} |C|^2}
\end{equation}
were $\alpha = 0.46 e^- sec^{-1}$ is the camera conversion rate, $R_{tr} = 87.5 \times 10^6  ~\text{counts} ~ s^{-1}$ the integrated counts for the trapped polariton condensate and $R_{pr} = 812 \times 10^6 ~\text{counts} ~ s^{-1}$ for the un-trapped condensate, $n_{tot} = T_{\text{optics}} \times   QE_{\text{camera}} =7.3 \times 10^{-6}$ is the total efficiency of the collection system: $T_{\text{optics}}$ is the efficiency of the optical setup, $QE_{\text{camera}}$ is the quantum efficiency of the camera. The polariton lifetime is $\tau = 10~ps$ and the photon fraction is $|C|^2 = 0.4$ The resulting steady state polariton numbers are:
\begin{gather*}
    N_{tr}= 150 \pm22\\
    N_{pr} = 1280 \pm 420
\end{gather*}
 Were $N_{tr}$ is the steady state number of trapped polaritons, and $N_{pr}$ is the steady state polariton count without an implanted trap, both at their respective $P = 1.6P_{th}$.

\renewcommand{\theequation}{C.\arabic{equation}}
\setcounter{equation}{0}
\section{Wave functions in reciprocal space}\label{AppA}

In the case of one-dimensional wave functions in quantum wires formed by effective potentials,
the reciprocal space emission profiles are given the Fourier transform of the real space wave function.
For even states the reciprocal space wave function reads
\begin{align}
    \tilde{\Psi}_+(k) = \frac{\sin((q-k)d)} {q-k} + \frac{\sin((q+k)d)} {q+k} + \nonumber \\
    \cos(qd) \left[ \frac{\exp(ikd)} {\varkappa - ik } + \frac{\exp(-ikd)}  {\varkappa + ik } \right]
\end{align}

For odd states the reciprocal space wave function reads
\begin{align}
    \tilde{\Psi}_-(k) = i\frac{\sin((q+k)d)}{q+k} - i\frac{\sin((q-k)d)}{q-k} +  \nonumber \\
    \sin(qd) \left[ \frac{\exp(-ikd)}{\varkappa + ik } - \frac{\exp(ikd)}{\varkappa - ik } \right]
\end{align}

In the case of radially symmetric wave functions in annular effective potential traps,
the reciprocal space emission profiles are given the 2D Fourier transform of the real space wave function (7) in the main text:
\begin{equation}
    \tilde{\Psi}_l (k,\varphi_k) = {1 \over 2 \pi} \int_0^\infty \rho d\rho \Psi_\rho(\rho) \int_0^{2\pi} e^{il\varphi} e^{-ik\rho\cos(\varphi-\varphi_k)}d\varphi
\end{equation}
Using the Hansen-Bessel formula and the Lommel integral, one can compute:
\begin{align}
    \tilde{\Psi}(k,\varphi) = -e^{il(\varphi_k-{\pi/2})} \left[ I_A(k) + I_B(k) + I_C(k) + I_D(k) \right], \\
    I_A = {A \rho_1 \over k^2 + \varkappa^2} \left[ k J_{l+1}(k\rho_1) I_l(\varkappa\rho_1) + \varkappa I_{l+1}(\varkappa\rho_1)J_l(k\rho_1) \right],\\
    I_B = {B \over k^2 - q^2} \left[ k \rho_2 J_{l+1}(k\rho_2) J_l(q\rho_2) - q \rho_2 J_{l+1}(q\rho_2)J_l(k\rho_2) - \right. \nonumber \\
    \left. - k\rho_1 J_{l+1}(k\rho_1)J_l(q\rho_1) + q\rho_1 J_{l+1}(q\rho_1)J_l(k\rho_1) \right],\\
    I_C = {C \over k^2 - q^2} \left[ k \rho_2 J_{l+1}(k\rho_2) Y_l(q\rho_2) - q \rho_2 Y_{l+1}(q\rho_2)J_l(k\rho_2) - \right. \nonumber \\
    \left. - k\rho_1 J_{l+1}(k\rho_1)Y_l(q\rho_1) + q\rho_1 Y_{l+1}(q\rho_1)J_l(k\rho_1) \right],\\
    I_D = -{D \rho_2 \over k^2 + \varkappa^2} \left[ k J_{l+1}(k\rho_2) K_l(\varkappa\rho_2) + \varkappa K_{l+1}(\varkappa\rho_2)J_l(k\rho_2) \right],
\end{align}

\begin{acknowledgments}
P.G.S. achlowledges support from by the Innovation Program for Quantum Science and Technology (2023ZD0300300), the Innovation Resource Allocation of Westlake University-Zhejiang Provincial Natural Science Foundation of China under Grant No. XHD24A2401, and the Westlake University project No. 102510036022301.\\
A.N. acknowledges support by the Russian Science Foundation under Grant No. 25-12-00135.\\
S.K.R. and H.O. acknowledge support from the Leverhulme Trust through grant No. RPG-2022-188.\\
We acknowledge E. Mavrotsoupakis for his contribution to the initial spectroscopic characterization of the samples. 
\end{acknowledgments}

\bibliographystyle{apsrev4-1}
\bibliography{apssamp}

\end{document}